\documentclass[nofootinbib,showpacs,aps,prc,reprint,floatfix]{revtex4-1}
        \usepackage{color}
\usepackage{epsfig}
\usepackage{graphicx}
\usepackage{amssymb}
\usepackage{amsmath}
\usepackage{upgreek}
\begin{document}
\title{Comment on "The relativistic Doppler effect: when a 
       zero-frequency\dots "}
\author{Z.\ Basrak$^1$}
\email{basrak@irb.hr}
\affiliation{$^1$Ru{d\llap{\raise 1.22ex\hbox
   {\vrule height 0.09ex width 0.2em}}\rlap{\raise 1.22ex\hbox
   {\vrule height 0.09ex width 0.06em}}}er
   Bo\v{s}kovi\'{c} Institute, Zagreb, Croatia}
\date{\today}
\begin{abstract}
  In the paper 
The relativistic Doppler effect: when a zero-frequency
shift or a red shift exists for sources approaching
the observer,
Ann. Phys. (Berlin) 523, No. 3, 239-246 (2011) / 
DOI 10.1002/andp.201000099 by C. Wang the use of an 
erroneous equation ended up at a number of faulty 
conclusions which are corrected in the present Comment. \\[-0.5em]

\end{abstract}
\maketitle

\noindent
In this Comment are corrected some results that are obtained 
in\ \cite{wang}. 
Instead of using the phase invariance and the time dilation 
in the derivation of the expressions for the Doppler shift 
and the aberration as in\ \cite{wang} we use the Lorentz
transformations (LT) of the four-dimensional (4D) wave vector. 
Otherwise, all notations are kept the same as in\ \cite{wang}.

Let us assume that the spectrograph is at rest in the 
laboratory inertial frame of reference (IFR) $S$.
The light source is at rest in the IFR $S'$ which is moving 
with velocity ${\mathbf v}$ relative to $S$ along the common 
$x, x'$--axis.
The components of the wave 4-vector in $S$ are 
$k^\mu = (\omega/c)(1,\cos\phi,\sin\phi,0)$ and in $S'$ they are 
$k'^\mu = (\omega'/c)(1,\cos\phi',\sin\phi',0)$ for which it 
holds $k^\mu k_\mu = k'^\mu k'_\mu = 0$.
Here, $\phi$ ($\phi'$) is the angle between the wave 
direction of propagation and $x$-- ($x'$--)axis, i.e.\ 
relative to ${\mathbf v}$.
The components of $k^\mu$ can be obtained by the LT of the 
components $k'^\mu$ which yields 
\begin{equation}
k^\mu = [\frac{\gamma\omega'}{c}(1+\beta\cos\phi'),
         \frac{\gamma\omega'}{c}(\cos\phi'+\beta),
               \frac{\omega'}{c}\sin\phi',0].
\label{keq}
\end{equation}
\noindent
Hence, the Doppler shift is given as 
\begin{equation}
\omega = \gamma\omega'(1+\beta\cos\phi') \\
\label{dseq}
\end{equation}
\noindent
whereas the equations that describe the change in the 
direction of wave propagation are 
\begin{equation}
\cos\phi = \frac{\cos\phi'+\beta}{1+\beta\cos\phi'}
\label{ceq}
\end{equation}
\noindent
and
\begin{equation}
\sin\phi = \frac{\sin\phi'}{\gamma(1+\beta\cos\phi')}.
\label{seq}
\end{equation}

In fact, if the observation of the unshifted line (i.e.\ of 
the frequency $\omega' = \omega_0$ from the atom at rest) is 
performed at an observation angle $\phi'$ in $S'$, the rest 
frame of the emitter, then the same light wave (from the same 
but now moving atom) will have the shifted frequency $\omega$ 
and will be seen at an observation angle $\phi$ (generally 
different from $\phi'$) in $S$, the rest frame of the 
spectrometer.
In astronomy the angular shift
\begin{equation}
\Delta = \phi'-\phi
\label{aeq}
\end{equation}
\noindent
is dubbed aberration.

\begin{figure}
  \includegraphics[width=80.mm]{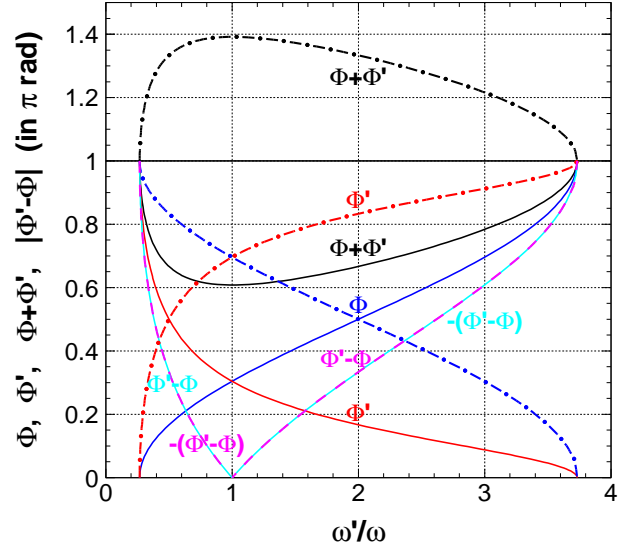}%
  \caption{\label{old_fg}
   (Color online.)
   Angles $\phi$ (blue), $\phi'$ (red), and $\phi+\phi'$ 
   (black) as well as the absolute value of the aberration 
   $|\phi'-\phi|$ as a function of $\omega'/\omega$.
   Results due to Eq.\ (\ref{weq}) are shown by the 
   full-line curves while those of Eq.\ (\ref{iweq}) by the 
   dash-dotted-line curves.
   The aberration, Eq.\ (\ref{aeq}) is shown by the 
   dashed-line curves, in magenta for IFR $S$ and in cyan 
   for IFR $S'$.
}
\end{figure}

The inverted relations between IFRs $S$ and $S'$ are
obtained by mere interchange of $\phi$ and $\phi'$
and $\beta$ by $-\beta$.
Thus, the inverted Doppler effect and cosine, Eqs.\
(\ref{dseq}) and (\ref{ceq}), read
\begin{equation}
\omega' = \gamma\omega(1-\beta\cos\phi)
\label{idseq}
\end{equation}
\noindent
and
\begin{equation}
\cos\phi' = \frac{\cos\phi-\beta}{1-\beta\cos\phi},
\label{iceq}
\end{equation}
\noindent
respectively.
We emphasize that Eqs.\ (\ref{idseq}) and (\ref{iceq}) have 
been derived by Einstein in his fundamental work on special 
relativity theory (SRT)\ \cite{ein905} and may also be found 
in many textbooks on SRT like e.g.\ \cite{pauli}.

\begin{figure}[t]
  \includegraphics[width=80.mm]{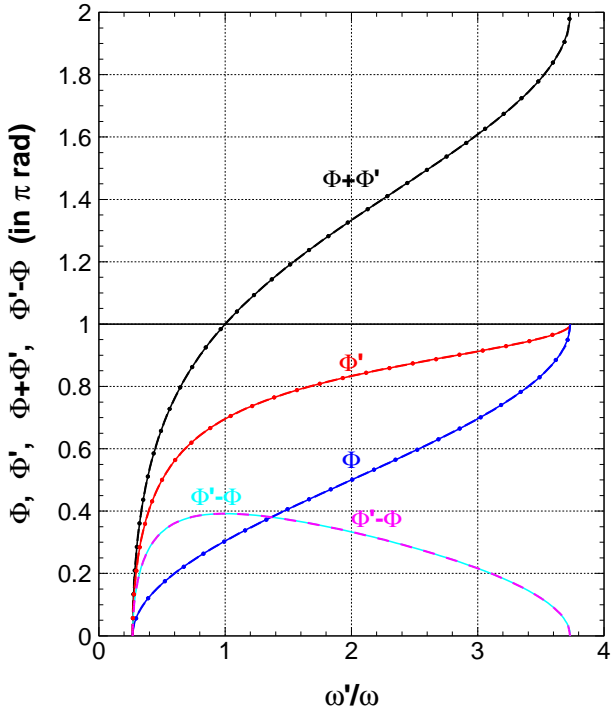}%
  \caption{\label{new_fg}
   (Color online.)
   Same as Fig.\ 1 but by using Eqs.\ (\ref{ceq}) and
   (\ref{iceq}) for $\phi$ and $\phi'$ as well as  
   (\ref{aeq}) for $\Delta$.
}
\end{figure}

In contrast to the above Eq.\ (\ref{iceq}) the cosine Eq.\ 
(6) in\ \cite{wang} reads
\begin{equation}
\cos\phi' = \frac{\beta-\cos\phi}{1-\beta\cos\phi}.
\label{weq}
\end{equation}
\noindent
This erroneous equation gives unphysical results that are
presented in Fig.\ 3 of Ref.\ \cite{wang} and repeated by
the full-line curves in Fiq.\ 1 here.
These curves pretend to describe the dependence of $\phi$,
$\phi'$, and $\phi+\phi'$ as a function of the ratio
$\omega'/\omega$ for $\gamma=2$.
Let us firstly examine the case of the light wave which is 
head-on approaching the observer, i.e.\ the case of 
$\phi = 0$ in IFR $S$.
Eq.\ (\ref{weq}) suggests an obviously unphysical behavior 
of such a wave because in the emitter IFR $S'$ it would 
have to recede with $\phi'=\pi$.
Equally unphysical is the case of a light wave which is 
receding from the observer at $\phi = \pi$ because it would 
have to move towards the observer in IFR $S'$ with $\phi' = 0$.
Further consequence of the erroneous Eq.\ (\ref{weq}) is 
that the Doppler zero-frequency shift (zfs), i.e.\ 
$\omega'=\omega$ occurs when the two position angles $\phi$ 
and $\phi'$ are equal or $\phi_{\rm zfs}=\phi_{\rm zfs}'$.
The author states \textit{\dots there is no frequency shift 
in such a case, although the light aberration must exist 
($\phi+\phi'\neq\pi$ for $\phi'=\phi$) \dots}\ \cite{wang}.
It is unclear from where the assertion that the light 
aberration vanishes for $\phi'+\phi=\pi$ does come?
In fact, it will be shown below (see Fig.\ 2) that for 
$\phi'+\phi=\pi$ the aberration $\Delta$ is at maximum.
Also, he claims \textit{the zero shift taking place at 
$\phi = \phi_{\rm zfs}$, where the aberration reaches a 
maximum}\ \cite{wang} which is an entirely contradictory 
statement because for $\phi = \phi'$ 
the aberration, Eq.\ (\ref{aeq}), vanishes.
For $\phi$ and $\phi'$ of Eq.\ (\ref{weq}) the resulting 
$\Delta$ is shown by the dashed curve.
Because $\Delta$ changes the sign for $\omega'>\omega$, 
in order to fit into the figure frame, in Fig.\ 1 is 
displayed the absolute value $|\Delta|$.

According to the principle of relativity the physical reality 
should not depend on the concrete IFR and coordinate basis 
used in describing it.
The most simple way to verify the correctness of an 
expression is to interchange the IFRs, i.e.\ $S$ and $S'$.
In that case Eq.\ (\ref{weq}) becomes 
\begin{equation}
\cos\phi = \frac{-\beta-\cos\phi'}{1+\beta\cos\phi'}
\label{iweq}
\end{equation}
\noindent
and its predictions for $\phi$, $\phi'$, and $\phi+\phi'$ 
are shown by the dash-dotted line curves in Fig.\ 1.
All three considered physical quantities $\phi$, $\phi'$, 
and $\phi+\phi'$ display an entirely different feature: 
these $\phi$ and $\phi'$ are mirror symmetric about $\pi/2$ 
relative to their previous graphs while their sum is 
symmetric about the angle $\pi$.
One again has $\phi_{\rm zfs} = \phi_{\rm zfs}'$ but its 
value is $\pi$ minus the previous one that was obtained 
from Eq.\ (\ref{weq}).
Although the aberration curve seems to be unchanged that is 
due to its absolute value.
Namely, with Eq.\ (\ref{iweq}) $\Delta>0$ for 
$\omega'>\omega$ and $\Delta<0$ for $\omega'<\omega$.

Figure 2 displays the correct observables $\phi$, $\phi'$, 
$\Delta$, and $\phi+\phi'$ obtained by using Eqs.\ 
(\ref{ceq}), (\ref{iceq}), and (\ref{aeq}), respectively.
Inverting the role of the IFRs $S$ and $S'$ gives the 
identical results as it should be (the full, Eq.\ 
(\ref{ceq}), and dash-dotted curves, Eq.\ (\ref{iceq}), 
are laying over each other).
Because $\phi$ and $\phi'$ are monotonically increasing 
functions of $\omega'/\omega$ such is also $\phi+\phi'$.
The aberration $\Delta$ is indeed maximal at 
$\omega'=\omega$ where $\phi+\phi'=\pi$ and 
$\phi_{\rm zfs}'=\pi-\phi_{\rm zfs}$.

Another way to verify the correctness of Eqs.\ (\ref{keq}) 
to (\ref{seq}) is to use a geometric approach to SRT from\ 
\cite{ive02l}.
As seen from Sec.\ 7.2 in\ \cite{ive02l} an abstract 
coordinate-free wave vector $k^{\rm a}$ is represented in 
$S$ ($S'$) by the coordinate-based geometric quantity (CBGQ) 
$k^\mu e_\mu$ ($k'^\mu e'_\mu$) comprising both the 
components $k^\mu$ and the 4D basis vectors $e_\mu$.
Any CBGQ is an invariant 4D quantity under the LT since the 
components transform by the LT and the basis vectors by the 
inverse LT leaving the whole CBGQ unchanged; it is the same 
physical quantity for relatively moving inertial observers.
It can be easily seen that with (\ref{keq}) it holds that 
$k^\mu e_\mu$=$k'^\mu e'_\mu$, which proves the validity of 
(\ref{keq}), i.e.\ of Eqs.\ (\ref{dseq}), (\ref{ceq}) and 
(\ref{seq}) and at the same time it disproves Eq.\ (\ref{weq}).

The author is indebted to Dr. T. Ivezi\'c for initiating 
this study, his constant encouragement, and the critical 
reading of the manuscript.
This work has been supported in part by Croatian Science
Foundation under the Project No. 7194 and in part by the
Scientific center of excellence for advance materials and
sensors.


\begin{thebibliography}{0}
  
\bibitem{wang}
\textsc{C.~Wang}, Ann.\ Phys.\ (Berlin) \textbf{523}, 239 (2011).
\bibitem{ein905}
\textsc{A.~Einstein}, Ann.\ Phys.\ (Leipzig) \textbf{17}, 891
(1905); in The Principle of Relativity (Methuen and Co.,
London, 1923), p.\,56.
\bibitem{pauli}
\textsc{W.~Pauli}, Theory of Relativity (Pergamon Press,
London and New York, 1958), p.\,19.
\bibitem{ive02l}
\textsc{T.~Ivezi\'c},
Foundations Phys.\ Lett.\ \textbf{75}, 27 (2002).
  
\end{thebibliography}
\end{document}